# Social Networks Research Aspects : A Vast and Fast Survey Focused on the Issue of Privacy in Social Network Sites

Mohammad Soryani

Mazandaran University of Science and Technology

Mazandaran, Iran

Behrooz Minaei

Iran University of Science & Technology

Tehran, Iran

*Abstract*— The increasing participation of people in online activities in recent years like content publishing, and having different kinds of relationships and interactions, along with the emergence of online social networks and people's extensive tendency toward them, have resulted in generation and availability of a huge amount of valuable information that has never been available before, and have introduced some new, attractive, varied, and useful research areas to researchers. In this paper we try to review some of the accomplished research on information of SNSs (Social Network Sites), and introduce some of the attractive applications that analyzing this information has. This will lead to the introduction of some new research areas to researchers. By reviewing the research in this area we will present a categorization of research topics about online social networks. This categorization includes seventeen research subtopics or subareas that will be introduced along with some of the accomplished research in these subareas. According to the consequences (slight, significant, and sometimes catastrophic) that revelation of personal and private information has, a research area that researchers have vastly investigated is privacy in online social networks. After an overview on different research subareas of SNSs, we will get more focused on the subarea of privacy protection in social networks, and introduce different aspects of it along with a categorization of these aspects.

*Keywords- Social Networks; Privacy; Privacy in Social Networks; SNS; Survey; Taxonomy;*

I. INTRODUCTION

In recent years several attractive and user-friendly facilities have been introduced to online society and we see an extensive and increasing participation of people in various online activities like several kinds of content publishing (blogging, writing reviews etc.) and having different kinds of relationships and interactions. The huge amount of information that is generated in this way by people has never been available before and is highly valuable from different points of views. An outstanding phenomenon that has had a significant influence on this extensive participation and includes a large part of generated information is SNSs (Social Network Sites). Maybe in past, to study about the relationships, behaviors, interactions, and properties of specific groups of people it was necessary to make a lot of effort to gain some not very detailed information about them, but in the new situation and with the emergence of online social networks, and the huge amount of various activities that are logged by their users, the desired information is accessed much more simple and with incomparably more details than before by researchers. This has led to different kinds of research with different goals which we will have an overview on in this paper. The benefits and stakeholders that may benefit from having this information or having the results of analyzing it are several but some of them are: commercial companies for advertising and promoting their products, sociologists to analyze the behavior and features of different societies, intelligence organizations to prevent and detect criminal activities, educational and cultural activists for promoting their goals, employers for acquiring information about job seekers, and generally any kind of information with any application that you may think of, related to people and human societies, may be obtained by having access to the information available on SNSs or the results of analyzing these information. In this paper we try to review some of the accomplished research on the available information of SNSs and present a categorization of research topics and subareas related to online social networks' information.

As a lot of peoples' published information is private and on the other side as we will see, having access to them has a lot of applications and benefits for different parties, letting them to be available with unlimited access has consequences that sometimes may be catastrophic. We will pay more attention to this issue in this paper.

In the following sections of this paper we will first introduce sixteen research subareas about online social networks while mentioning a few of accomplished studies related with them. After that we will have a more focused review on another important research subarea namely privacy protection in social networks, and will present a categorization of its different aspects. We will conclude at the end. To have a look at the whole picture of the categorization from above, Fig.1 shows several research sub-topics about SNSs and Fig.2 extends the topic of privacy and presents a categorization of several aspects of privacy in social networks.

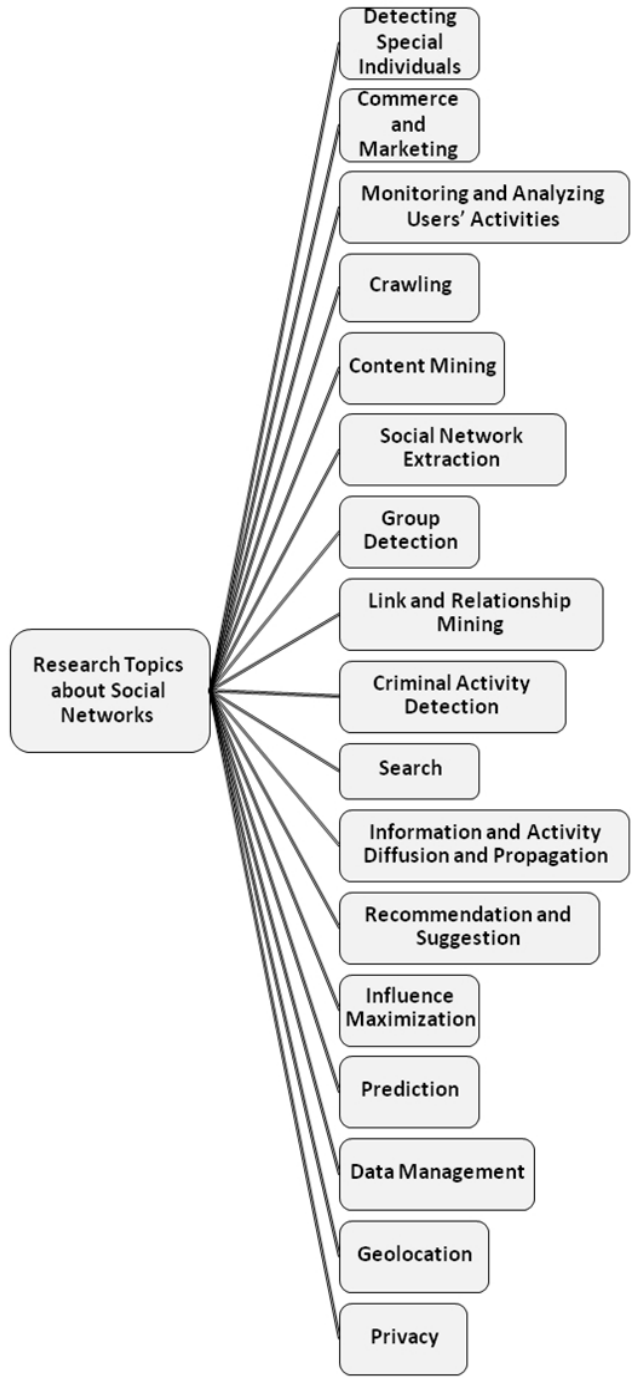

Figure 1. A categorization of research topics in social networks area.

## II. A CATEGORIZATION OF RESEARCH TOPICS IN SOCIAL NETWORKS AREA

**1) Detecting Special Individuals:** Some people with special characteristics may be attractive for some companies, manufacturers, organizations, etc. for example it may be desirable to find special persons with high skill in a special field or to find most influential persons in propagating some special kind of content. In [3] a social search engine named Aardvark (http://vark.com) is presented that needs to find the best person for answering a specific question, and one of its information resources is people's profiles on facebook. In [4] some work is done towards forming a team of experts from members of a social network. In [5] to specify influential persons within Twitter, ranking people based on their followers, PageRank and number of retweets is investigated. In [6] some definitions are defined for different people whose actions impacts on making the same actions by others and such people are called leaders. In this paper some algorithms are presented for detecting these people by the use of a social graph and a table which contains users' actions. In [14] some references are cited in which some methods for extracting most important (central) members are presented. It has mentioned strengths and weaknesses of some metrics. In [21] Content Power Users (CPUs) in blog networks are defined as users whose published content has a lot of impact on other users' actions. In this paper a method for identifying these users is presented and some other research works about detecting highly influential people in social networks are cited.

**2) Commerce and Marketing:** Advertisement in SNSs can be targeted [11] [31] [6]. Targeting users whose activity influences others could be beneficial for companies [6]. As is mentioned in [11] a manufacturer can select a number of users and give them its product with some discount or even free, and hope that their influence on other users promotes their product. In the case of discounting the amount of discount is exposed to discussion. In [31] using users' profile information and the information about their activities towards targeted advertisement is mentioned.

**3) Monitoring and Analyzing Users' Activities:** In [12] users' behavior in a social network is analyzed to identify the patterns of closeness between colleagues. Paper [24] notes the importance of awareness about users' participation patterns in knowledge-sharing social networks for researchers and social network industry; and analyses users' activities in three social networks. Some results that are different from common assumptions are reported.

**4) Crawling:** To analyze the information of SNSs first we must acquire it. One of the most important ways to do this is using crawlers. Crawlers generally should have some specifications like being up to date and having mechanisms to prevent fetching the same page more than once. According to special characteristics of social networks like the huge size, and the auto crawling prevention mechanisms that SNSs use it seems that we need special kind of crawlers. In [10] noting the large size of data, and the different way of data presentation in social networks a parallel crawler is proposed for crawling social networks. In [39] [7] facebook is crawled and the problem of facing with CAPTCHAs is noted. In [38] the ethicality of web crawlers is discussed.

**5) Content Mining:** One kind of information that is made by users is the content that they put in the sites in different ways. In [8] according to the real-time nature of twitter, an algorithm for monitoring and analyzing the tweets is proposed towards detecting a specific event. In this work a system is implemented to detect earthquakes in Japan and is able to do so with high accuracy. In [9] new type of texts that are published on SNSs and are usually short with an informal form of writing (called social snippets) are investigated and some applications of analyzing such texts are mentioned. Their focus is on keyword extraction from this kind of texts. In [36] mentioning the applications of identifying the quality of users' reviews about different issues, using social networks information to improve reviews quality identification is discussed. In [22] the necessity of applying automated language analysis techniques towards security in digital communities including social networks is mentioned and an approach is proposed for detecting a special kind of malicious activity.

**6) Social Network Extraction:** There are a lot of data in various forms on the web that apparently do not have the structure of a social network but with some mining activities on them like extracting the identity of data owners and the relations between them it is possible to extract the social network that relies beneath these data. Examples of such studies are [1] and [2]. In [1] the information of a message board is used and in [2] a system named ArnetMiner [http://www.arnetminer.org] is presented that extracts a social network of researchers. In [21] extraction of social network using a blog is mentioned.

**7) Group Detection:** In [4] identifying a group of skillful people to accomplish a specific job with a minimum communication cost in a social network is discussed. In [25] an efficient algorithm for large social networks named ComTector is presented for detecting communities. In [33] grouping in a social network is done towards detecting the backbone of a social network.

**8) Link and Relationship Mining:** In [12] relationship closeness is investigated based on the behavior of users in a social network inside a company. It is mentioned that Some behaviors are a sign of professional closeness and some are a sign of personal closeness. In [13] an approach is presented for estimation of relationship strength, and is evaluated with facebook and linkedin data. In [34] an approach for inferring the links that exist but are not observed is presented and a good survey about link mining in social networks is cited.

**9) Criminal Activity Detection:** Some specifications of SNSs like presence of great number of various people and new ways of communication has attracted criminals to use them for their malicious activities, so to prevent and detect such activities some special research is necessary to be done. In [22] some challenges about law enforcement and the necessity of using automated language analysis techniques for active policing in digital communities is mentioned. It notes some applications of using these techniques like identifying the child predators who pretend to be a child. In [29] a system is proposed for identifying suspects with the help of social network analysis (SNA). In [16] the application of SNA in criminal investigation and yet protecting privacy at the same time is discussed. In [32] the importance of clustering web opinions from intelligence and security informatics point of view along with some criminal activities that could be done in this space are mentioned and a clustering algorithm for detecting the context of the discussions available at social networks is presented.

**10) Search:** Search engines, both general purpose and special purpose ones use different information as parameters for ranking their search results. SNSs may be a valuable source of information to improve the ranking. For example special characteristics of people acquired from social networks, may be considered for ranking search results tailored to each individual's characteristics resulting in personalized search. In [19] using the structure of a social network toward improved result ranking in profile search is studied and using the social graph for improving ranking in other types of online search is mentioned as a future research. In [3] a search engine is introduced that instead of looking for appropriate documents related with the given query, it looks for suitable persons for answering the given question. To do that it gathers information about people from different resources including SNSs. In [37] towards leveraging the information about searchers in social networks for document ranking, a framework named SNDocRank is presented. Also a study about personalizing search results using users' information is cited.

**11) Information and Activity Diffusion and Propagation:** The way that information and activities propagate through a social network is another area that is worth to investigate and studying about it can have various benefits. For example commercial companies may be interested in the results of such studies to improve the spread of information about their products, or educational and cultural activists may benefit from it for promoting their goals. In [20] information propagation in blogs is studied and some applications for such a study are mentioned. In [6] the spread and prevalence of users' actions is investigated over time, to identify users whose actions have influence on other users' actions.

**12) Recommendation and Suggestion:** According to [27] the goal of a recommender system is to recommend a set of items to a user whose favorite items are similar to them. In this work some algorithms have been designed and implemented for such systems in social networks. In [28] some techniques are presented for link prediction and the application of such techniques in friend recommendation in social networks is mentioned. In [23] some important aspects of research related with social recommendation that could be done are mentioned.

**13) Influence Maximization:** The problem here is to find a number of persons whose scope of influence is maximum; for example a company that has developed an application for a social network and wants to market it on that social network and can afford to invest on limited number of users (for example for giving gifts to them) would like to choose these users so that the extent of final influenced users is maximum [35]. In [26] assigning roles to users and application of being aware of these roles in influence maximization is mentioned. In [35] in addition to improving another algorithm called greedy, some heuristics are presented that run much faster.

**14) Prediction:** By studying the information of SNSs it is possible to predict some events that may happen in future. For example some research has been done recently to predict future links in social networks [34]. In [28] some techniques for predicting the links that may be established in future are presented.

**15) Data Management:** Managing the huge volume of SNSs' data has several aspects and due to special features of this kind of data, needs specific research. For example the structure of storing data is very important and can affect the amount of needed storage. In [30] a study is done about compressing social networks and a new method for compressing social networks is presented. Some of the similarities and differences between this problem and the problem of compressing web graph are found. Their results show vast difference between compressibility characteristics of social networks and web graph.

**16) Geolocation:** Detecting the location of the user can have several applications including personalization. Among the studies in which SNSs are used to detect the location of users is [18]. In this work researchers of facebook have mentioned some applications of knowing the location of users like news personalization, and with pointing to inconsistency of results when using ip address for geolocation they have presented a method for detecting the location of the user using information about the location of her friends. It is also mentioned that their algorithms could be run iteratively towards identifying the location of most users who have not provided any information about their location.

**17) Privacy:** As the focus of this paper is on the area of privacy in SNSs, we will present a broader overview on several aspects of this area along with introducing a taxonomy of these aspects which is shown in Fig.2.

**17-1 Defence**
**17-1-1 Helping Users**
Users need to be informed about the consequences of their various information publication and their activities; need to know which part of their information is accessible and for whom; need to have some facilities to control the way that people can access their information, and need to get all of these requirements in a simple and understandable way that does not need a lot of time and effort. A user gets acquainted with the matter of privacy from the first steps of her experience on SNSs by facing with privacy policies text. Definitely you too have faced with such a text and confirm that they are not very pleasant for users and a lot of people accept and pass over them without reading. In [42] the challenge of showing users' privacy related issues to them in an understandable way is presented with an interesting example where a possible interpretation of a privacy-related text may be different from what really happens. In a study that is done on six well-known SNSs [51], it is mentioned that privacy policies often have internal inconsistencies and also there is a lack of clear phrases about data retention.

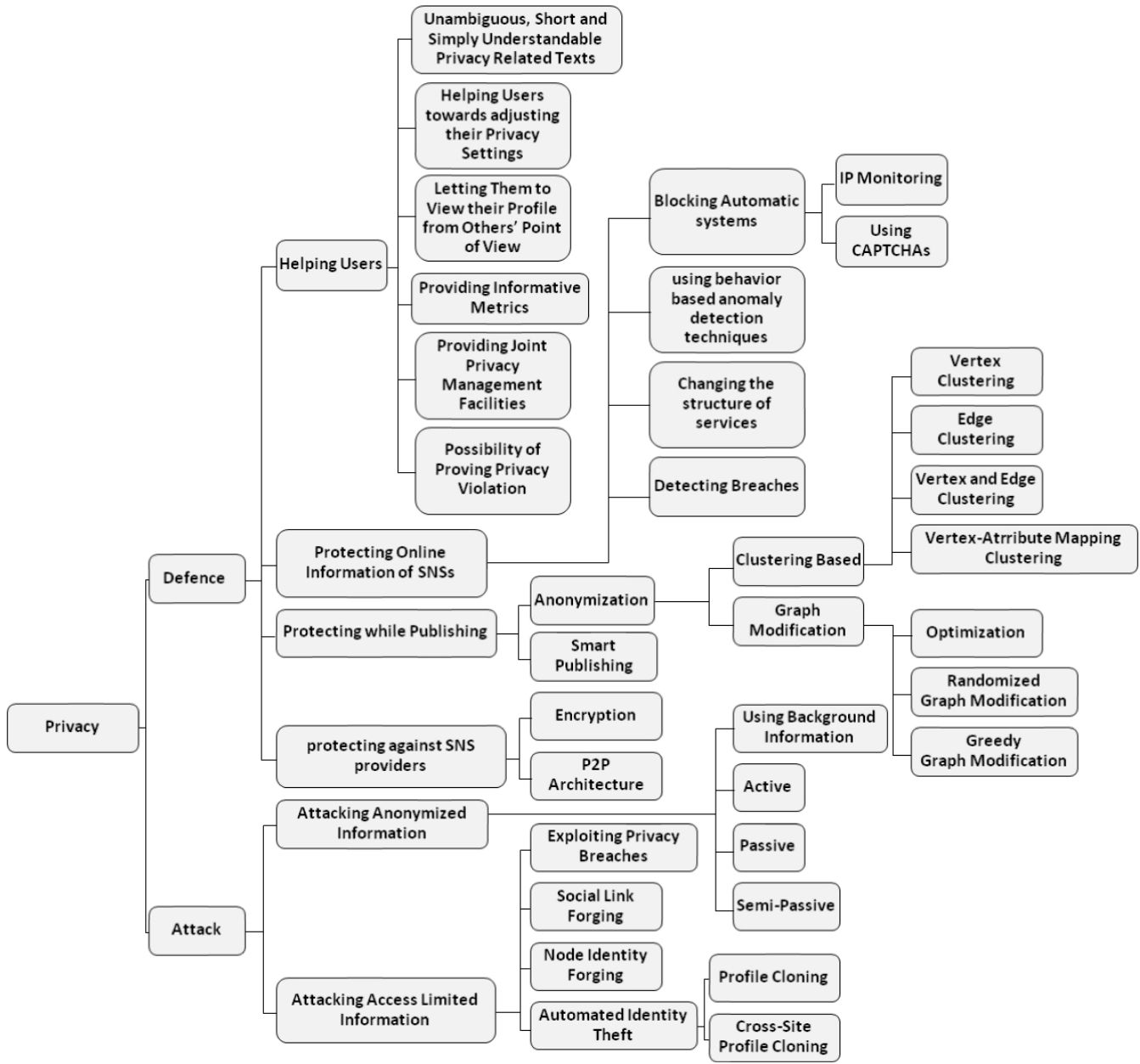

Figure 2.   A categorization of privacy related aspects in Social Network Sites.

Overall privacy related texts like privacy policies should be unambiguous, as short as possible, and simply understandable. This can be considered as the first step of helping users towards their privacy protection. After accepting the privacy policy agreement, there are some flexible privacy settings that could be adjusted and help the user to specify the way her information is accessible. Since it is difficult for the average user to adjust these fine grained and detailed settings [44], some research towards helping users to do that in the best way could be useful. In [44] a work has been done toward automatically adjusting these settings with the minimum effort of the user. In this study a wizard is introduced that builds a classifier based on the user's answers to the systems requests and uses it to automatically adjust the settings. The possibility of using a limited amount of user's input to build a machine learning model that concisely describes the privacy preferences of a user, and using this model for automatically adjusting privacy settings of a user is presented in this paper. Some other interesting points about this approach are the adoption of the system after a new friend is added by a user (it presents some information about the user's preferences), and the possibility of viewing and modifying the obtained model by advanced users. Another helpful facility (the work is presented by Lipford et al. and is cited in [44]) is a tool that makes the user able to view her own profile from the view point of each of her friends and consequently observing the results of her settings. It seems that some sort of this approach is adopted by facebook [44]. In [54] a method for suggesting privacy settings to newcomers is presented and the importance of these primitive settings due to the users' tendency towards keeping them is noted. In this study to some extent, a review about how to use machine learning to prepare and suggest primitive privacy settings that are more probably useful for users is presented. Another helpful aid for users' better privacy protection is providing some informative metrics by which users can obtain concise useful information. In [42] it is suggested to use an approach which uses data mining and potentially other AI tools to find out the amount of difficulty that accessing users' information has, and to provide some metrics for showing that, that is a useful tool for informing users about their privacy risk. Another metric that is presented in [65] is "privacy score". In this study some models and algorithms have been presented for calculating this metric and some mathematical models are developed to estimate the sensitivity and visibility of the information, which are influential on privacy risk. Sometimes users do not behave appropriately regarding their privacy while using SNSs. Understanding these risky behaviors, their causes, and then informing people and developing preventive mechanisms could be another way of helping users. In [47] a few of these risky behaviors are mentioned : incaution in accepting friendship requests, clicking on links received from others without enough caution, reacting to suspicious friendship requests after accepting them (not before) and therefore letting suspicious users to access information, interaction with fake profiles and overall, high implicit trust that exists in SNSs. A study is cited there in which 41% of 200 users whom a friendship request were randomly sent to, accepted it, and most of the users had not limited access to their personal profile information. In another study which is also cited in [47], publicly available information of some people has been gathered from some SNSs, and have been used in phishing emails; results show that targeted people whose received emails contains some information about them or their friends are more likely to get involved. In [46] [50] and [59] some surveys have been conducted as a way for studying users and acquiring information that leads to better helping them. In [46] interviews and surveys are used to investigate the effective factors that cause personal information exposure by students on facebook and why they do it despite the existing concerns. How to defend them against privacy threats is also studied. In [59] a survey is conducted in which the participants are highly educated and some information about users' behaviors, viewpoints, and concerns about privacy related issues are gained. Notwithstanding all of the existing threats and risks about private information disclosure, however users extensively tend to engage in social relations and interactions in SNSs and naturally each of these relations and interactions needs some information exposure. In [40] a study is done with the goal of determining the least information that needs to be shared to accomplish a specific interaction. An interesting study towards preparing helpful tools for users to protect their privacy is [53] in which helping users to jointly manage the privacy of their shared content is studied. Sometimes people who benefit from or get harmed by publication of a specific information unit are multiple, and publishing such information may put the privacy of several people at risk. An example of this kind of information that is extensively being published over SNSs is a photo in which several people exist. Photos are sometimes tagged and/or some additional information about them is available besides them. Support for common ownership in SNSs, and the requirements of solutions namely being fair, lightweight, and practical are mentioned as some issues that exist about joint management of privacy in [53]. In [59] it's been tried to find a way for a user to express her privacy preferences, and a method to do this with the aim of being easy to understand by other users and also being machine readable (so that it can be used by other service providers and third parties) is presented. It is mentioned that they intend to use cryptographic techniques to provide the possibility of proving privacy violation for the data owner (for example to a legal authority) which is another helpful facility towards users' privacy protection.

**17-1-2  Protecting Online Information of SNSs**

Let's assume that we have given all of the helpful information to users to properly manage their information publishing and protecting their privacy. Besides that we have provided them with best tools for adjusting their privacy settings in a fast, accurate, and simple way. Is it the time to relax and feel like we have done all we could do, or there are other issues that we should take care of and study about? The problem is that some of those who are interested in having users' information do not give up and try to exploit from any possible way (sometimes legal and sometimes illegal) to acquire their desired information. In addition to studies that have been done towards directly helping users, it is necessary to make some efforts to keep the online information out of undesired reach. In this part we will take a look at this issue. A study on five Russian SNSs [50] shows that despite more concerned users, there is a significant gap between their providers and western SNS Providers about understanding the privacy related concepts and preparing appropriate defencive mechanisms; and overall the privacy condition is reported to be catastrophic that leads to exposure of a large amount of users' information. An important tool that is used to acquire the online information of SNSs is a bot, with which intruders are able to accomplish their actions automatically and with a large scale. Among the actions that could be done with these automatic systems are: crawling, creating fake profiles, and sending forged friendship requests. Obviously something needs to be done to stop these systems. In fact SNS providers should somehow determine whether the requests are sent by a real person or by an automatic system. One thing that could be done is to monitor IP addresses. Another method that is common for online automatic activity detection is using CAPTCHAs [http://www.captcha.net, 47]. A CAPTCHA is a program that protects websites against bots by creating tests that human can pass but programs cannot [http://www.captcha.net]. According to [47], facebook uses an adoption of reCAPTCHA (a related reference is cited) solution which is developed at Carnegie Mellon university. Automatic systems may have mechanisms to pass CAPTCHAs. Also they can exploit the possibility of changing CAPTCHAs that some websites offer, to find a CAPTCHA which they are able to pass. To deal with this, the rate of presented CAPTCHAs could be limited. [47]

In [50] it is concluded that most Russian SNSs do not prevent automatic profile crawling appropriately. Also according to [47] in some cases there is a lot of improvement possibility to make CAPTCHAs more difficult to break, and not every SNSs try enough to make automatic crawling and access more difficult. Another way of prevention and detection of suspicious activities is using behavior based anomaly detection techniques that can reduce the speed of the attack and its economical feasibility [47]. Although SNSs should try to protect the privacy of users and keep their information accessible only in the way that they have determined by their settings, however there may be some privacy breaches. In [47] an interesting example is mentioned. It is said that according to similar characteristics of SNSs, an extendable model could be built and potential breaches could be detected by formal analysis of this model. Changing the structure of services may help to increase the privacy protection level for users. For example in [67] extending link types is mentioned as a helpful action for privacy protection. It means instead of simply just being connected or not connected, people could for example specify the direction of their connection or the amount of trust which exists in the relationship. An example of its benefit is when a malicious user succeeds to fool another user and establish a connection with her. In this situation using trust degree can decrease the consequences of this connection. [67]

A kind of extended link types is recently adopted by google's SNS [http://plus.google.com].

In [59] and [60] using cryptographic techniques is proposed to prove privacy violations, that can prevent unauthorized access, and if happened compensate or decrease the damage.

### 17-1-3 Protecting while Publishing
#### 17-1-3-1 Anonymization

So far we have discussed privacy protection by means of limiting access to users' personal information. In this section we are going to take a look at a kind of protection that aims to protect the privacy and publish information at the same time. As we mentioned, there is a high interest and desire with various motivations to have access to SNSs data. A method that is used to publish this data and protect users' privacy at the same time is called anonymization. An informal definition of anonymization in the context of privacy protection is replacing information that its revelation may damage users' privacy (email, address etc.) with other harmless data. In [36] and [52] the tradeoff between privacy and utility of anonymized data is discussed. A good survey about anonymizing social network information is [17]. In this study anonymization methods have been categorized and according to it, the state-of-the-art methods for social network information anonymization are clustering based approaches and graph modification approaches. Clustering based approaches include four subcategories of vertex clustering, edge clustering, vertex and edge clustering, and vertex-attribute mapping clustering. Graph modification approaches include three subcategories of optimization, randomized graph modification, and greedy graph modification. In [43] a weakness of past studies about anonymization is mentioned, which is their focus on methods that consider a single instance of a network, while SNSs evolutionally change and the information that does not reflect these changes is not enough for every analysis. In this paper some studies that has been done about the evolution of social networks are mentioned and also it is noted that anonymization of different instances taken in different times is not sufficient and comparing them leads to information revelation. Researchers of this paper have cited a report of their own in which they have proposed an approach for this problem. As we will see in the attacks section the beneficiaries still try to acquire their desired information and try to extract it even from anonymized data, so some studies towards improving anonymization techniques and overcoming their weaknesses like [69] have been and will be done. Researchers in [48] believe that in the area of users' privacy, mathematical rigor is needed towards having clear definitions about the meaning of comprising privacy and the information that the adversary has access to.

#### 17-1-3-2 Smart Publishing

An application of Social Network Analysis (SNA) is criminal investigation [16] but it seems to somehow have contradiction with the matter of users' privacy. An interesting approach is presented in [16] with which without direct access to the SNS information and even their anonymized form, only some results of SNA (in form of two centrality metrics) is given to investigators and gives them the opportunity to send queries without privacy violation.

### 17-1-4 Protecting Against SNS Providers

Another privacy related concern is about inappropriate or undesired use of users' information by SNS providers

[67] [59] [63]. Towards solving this problem a key management scheme is presented in [15]. In the proposed method, information is encrypted and SNS providers are unaware of the keys. It is claimed that it does not have a weakness of some other related works and does not require the user and her information viewer to be present at the same time. In [55] a client-server based arichitecture is proposed for protecting users' information from SNS providers' access in which information is transferred as encrypted blocks. In [67] peer to peer (P2P) architecture for SNSs and researchers tendency for designing this architecture for next generation of SNSs is mentioned. Some of the advantages of client-server architecture over P2P architecture (like more efficient data mining in a central repository), and shortcomings of using client-server architecture with encryption (like some relations of users being detectable by other data like IP addresses) are also noted. A combination of P2P architecture and a good encryption scheme is noted as a better solution for privacy protection in SNSs.

### 17-2 Attack

As it was mentioned there are interested applicants that do not give up after we put some barriers on their way and try to make unauthorized information out of their reach. They still try to attack and pass the barriers and reach their desired information. In this part we take a look at these attacks from two points of view, attacks on anonymized data and attacks on access limited information.

### 17-2-1 Attacking anonymized Information

Despite the efforts towards protecting users' private information when publishing SNSs information using anonymization, this information is still exposed to a special kind of attacks that aim at discovering information in anonymized information. Adversaries may use some background information to accomplish this kind of attacks. [17] [52] [48]

For example the attacker may have some specific information about her target (the person who the attacker intends to get access to her information), and be able to recognize her target among the anonymized information [17]. In [48] three types of attacks are presented as active attacks, passive attacks, and semi-passive attacks : Passive attacks are described as those in which attackers begin their work to detect the identity of nodes after anonymized information is published; at the other side in active attacks the adversary tries to create some accounts in the SNS and establish some links in the network so that these links will be present in the anonymized version of the information; in semi-passive attacks there is no new account creation but some links are established with the target user before the information is anonymized. Having background information (both the information that the attackers themselves has injected to the network and the information that they have acquired in other ways) is an important tool in the hand of anonymized information attackers. In [17] some examples of background information are mentioned like attributes of vertices, vertex degree, and neighborhoods. In [69] using neighborhood information is presented as an example of a type of attacks called neighborhood attacks. In [52] vertex degree is said to be the most vastly used parameter. In [48] some studies about using interesting information like user prepared text for attacking anonymized data are cited (although in different contexts from the SN context of that paper). In [39] good information is presented about anonymization and deanonymization. In this paper a passive method is proposed for identity detection in anonymized information, and using it along with the information of two well known SNSs (twitter, flickr) notable results are obtained regarding identity detection of some members of these SNSs in the anonymized graph of twitter. In this work the information inside flickr is used as background information. An important point mentioned in [43] is that in the area of social networks anonymization, the main focus so far has been on a single instance of the network's information in a specific time, and this is inconsistent with the very dynamic nature of these networks. It is noted there that having several copies of anonymized data of a social network, which are taken in different times may lead to information revelation by comparing these copies.

### 17-2-2 Attacking Access Limited Information

Using the facilities which users are given, to adjust how their information could be accessed, they can make their information not to be accessible by everyone, and specify different limitations for different parts of their information, for example a person may set her pictures to only be viewable by her close friends. Attackers certainly would try to cross these borders. In [67] "social link forging attacks" and "node identity forging attacks" are mentioned. The former means deceiving a user and convincing her to establish a link (may include impersonating one of her friends by the attacker), the latter means creating several fake identities and pretending to have several personalities in a SNS. In [42] a breach is detected in linkedIn, and using the presented method the contacts of a victim are extracted. Increasing the credibility of phishers' messages using the credibility of people who are connected with the victim is noted as a motivation for this kind of attacks. Some more complicated attacks are also cited in this paper. In [47] a type of attack called "automated identity theft" along with its two subtypes called "profile cloning" and "cross-site profile cloning" is presented. In these attacks the attacker tries to make fake profiles which appear to belong to persons who really exist and have profiles either in the target SNS or other SNSs. In this study a prototype of an attack system for performing attacks is presented which performs crawling, users' information gathering, profile creation, message sending, and tries to break CAPTCHAs. Some experiments have been done on five social networks including facebook and linkedin.

III. CONCLUSIONS

Emergence of social networks and increasing participation of people in activities in these sites along with the huge amount of various information like interactions, reviews, interests and different kinds of published contents that are logged by users have attracted researchers and other parties to have access to this information or to the results of analyzing it. This information has never been available with such a huge volume, detail, and ease and speed of access before. A few number of those who are interested in having this information or the results of analyzing it alongside their motivations are: commercial companies for advertising and promoting their products, sociologists for analyzing the behavior and features of different societies, intelligence organizations for preventing and detecting criminal activities, educational and cultural activists for promoting their goals, and employers for acquiring information about job seekers. In this paper along with introducing some of the studies in this area, a categorization of research subareas was presented and a base has been provided for researchers to briefly get acquainted with some new, attractive and useful research areas. A categorization of reviewed research subtopics is illustrated in Fig.1.

People always have some private information that do not want to be exposed to public access, and if accessed by some adversaries, may have some (sometimes catastrophic) consequences. So in this survey we focused on the issue of protecting users' privacy and presented a categorization of different aspects of this area. This categorization is illustrated in Fig.2.